\documentclass[
 reprint,
 amsmath,amssymb,
 aps,
 prl,
 longbibliography
]{revtex4-1}

\usepackage{graphicx}
\usepackage{dcolumn}
\usepackage{bm}

\def\beq{\begin{equation}}
\def\ee{\end{equation}}

\def\inn{_\mathrm{in}}
\def\out{_\mathrm{out}}

\newcommand{\set}[1]{\{{#1}\}}
\newcommand{\dd}[1]{\mathrm{d}{#1}}

\newcommand{\period}{\mathcal{T}}
\newcommand{\dfrequ}{\Omega}

\newcommand{\dai}{\dot{a}_i(t)}
\newcommand{\sv}{A_i(t)}

\newcommand{\dmij}{\dot{n}_{ij}}
\newcommand{\mij}{n_{ij}}
\newcommand{\occi}{\delta_{i(t),i}}

\newcommand{\expval}[1]{\langle{#1}\rangle}

\newcommand{\pst}[3]{p^\mathrm{ps}_{#1}(#3; #2)}
\newcommand{\jst}[2]{j^\mathrm{ps}_{ij}(#2; #1)}
\newcommand{\tst}[2]{t^\mathrm{ps}_{ij}(#2; #1)}
\newcommand{\dfrequaux}{\tilde{\dfrequ}}

\newcommand{\gbound}{\mathcal{S}}

\newcommand{\act}{\mathcal{A}}
\newcommand{\bound}{C}

\begin{document}
\preprint{APS/123-QED}
\title{Operationally accessible bounds on fluctuations
	and entropy production
	in periodically driven systems\\~\\}

\author{Timur Koyuk and Udo Seifert\\~
}
\affiliation{
{II.} Institut f\"ur Theoretische Physik, Universit\"at Stuttgart,
  70550 Stuttgart, Germany
}
\date{\today}

\begin{abstract}
For periodically driven systems,
we derive a family of inequalities that relate entropy production with experimentally
accessible data for the mean, its dependence on driving frequency, and the variance of a large class of observables.
With one of these relations, overall entropy production can be bounded by just observing
the time spent in a set of states. Among further consequences, the thermodynamic efficiency both of isothermal cyclic engines like molecular motors under a periodic load
and of cyclic heat engines can be bounded using  experimental data without requiring
knowledge of the specific interactions within the system. We illustrate these
results for a driven three-level system and for a colloidal Stirling engine. 
\end{abstract}

\maketitle

Periodically driven open systems typically reach a periodic steady state since the coupling to the environment prevents unlimited heating up. For meso- and nanoscopic systems, such a
steady state can still exhibit significant fluctuations~\cite{jung93}.
Beyond the quantum domain, in which such systems have found considerable attention recently, see, e.g.,~\cite{gasp14,shir16a,hart17,rest18,wang18,gamb19} and refs. therein, colloidal systems provide a major paradigm \cite{cili17}. Likewise, chemical and
biophysical systems on the molecular and cellular scale that are
subjected to periodic mechanical, optical or chemical stimuli fall into this wide class, see, e.g.,~\cite{haya12a,erba15}. Heat engines and
cooling devices coupled cyclically to baths of different temperature provide a further paradigm \cite{schm08,espo10,blic12,abah12,zhan14,verl14,camp15,peko15,uzdi15,bran15b,rossn16,kosl17,mart17}, 

One obvious question for any such system is whether or not the entropy production associated with the periodic driving can be inferred, or, at least, bounded, using only experimentally accessible observables without having detailed knowledge of the interactions or the internal structure of the system. This question is thus a non-trivial one whenever power input and power output are not both directly measurable, which is the case when not all degrees
of freedom that couple to these currents are observable. This situation
is \textsl{inter alia} typical for all systems undergoing chemical reactions. Likewise, it holds for systems driven by electric fields since
one cannot observe the motion of all charges, in particular, in (soft) condensed matter systems. 

For the arguably simpler class of non-equilibrium steady states
in systems under constant, time-independent driving, a
universal, experimentally accessible relation has recently
been found that achieves this aim for classical systems.
The so-called thermodynamic uncertainty relation (TUR)~\cite{bara15,ging16}
\beq 
\sigma D_J/J^2\geq 1
\label{eq:intro:TUR}
\ee
relates the entropy production rate $\sigma$ with the mean value of 
any current $J$ and its variance, or dispersion, $D_J$,
all defined
more precisely below.
Besides stating a bound on the
typically not directly accessible entropy production, this relation can also be interpreted as a bound on the precision of a process
in the sense that small fluctuations, i.e., high precision or low
uncertainty, requires a minimum amount of entropy production.
As specific applications, the efficiency
of molecular motors can be bounded using only experimental data~\cite{piet16b,hwan18} and design principles for self-assembly can be derived~\cite{nguy15}. For steady-state heat
engines, the relation shows that an inevitable side-effect of reaching Carnot efficiency
at finite power are diverging power fluctuations~\cite{shir16,camp16,pole17,piet17a,holu18}. Refinements
and generalizations of the TUR have been found for diffusive dynamics~\cite{piet15,ging16a,pole16}, for data allocated over a finite time~\cite{piet17,horo17,dech17,pigo17},
for 
ballistic transport~\cite{bran18}, for underdamped Langevin dynamics with and
without magnetic fields~\cite{fisc18,dech18,chun19}. Rather than looking at the fluctuations of currents, it is also possible to constrain the fluctuations
of time-symmetric quantities like residence times or activity~\cite{nard17a,maes17,terl18} and the fluctuations of first passage times~\cite{ging17,garr17}. While a few quantum systems
have been investigated, a systematic picture in the quantum realm is still missing~\cite{maci18,agar18,carr19,ptas18,guar19,bran18}.

An early counter-example has shown that a naive extension
of the thermodynamic uncertainty relation from steady-state systems to periodically driven ones is not admissible~\cite{bara16}.
Subsequent attempts to find an analog for periodically driven systems 
comprise Proesman and van den Broeck's bound valid for \textsl{time-symmetric}
driving that, however, for a small frequency of driving leads to a rather weak bound~\cite{proe17}. 
Barato \textsl{et al} replace $\sigma$ in eq.~\eqref{eq:intro:TUR} by a modified entropy production rate
that requires detailed knowledge of dynamical properties of the whole system~\cite{bara18b}. 
In a follow up~\cite{bara18c}, a whole class of
such modified entropy production rates were discussed
that, from an operational perspective, are arguably not
that useful since they
require input that is typically not available in experiments. 
The same holds for our generalization introducing an effective entropy production for further
types of currents~\cite{koyu19} and for a further scheme~\cite{rots16}.
Recently, Proesmans and Horowitz introduced a modified uncertainty relation for hysteretic currents that overcomes some of the above mentioned short-comings~\cite{proe19}.
However, from the perspective of thermodynamic inference~\cite{seif19}, an operationally accessible relation, which allows to bound entropy production and which becomes the TUR for steady-state systems, is still missing for periodically driven systems.

As a main result of this Letter, for systems driven with a 
period $\period\equiv2\pi/\dfrequ$, 
we will derive a family of universal bounds that relate, \textsl{inter alia}, the entropy production with fluctuations of observables. Applied to
 current fluctuations, we get specifically
 \beq
\sigma(\dfrequ)D_J(\dfrequ)/J(\dfrequ)^2\geq [1-\dfrequ J'(\dfrequ)/J(\dfrequ)]^2 .
\label{eq:main_result_J}
\ee
The left-hand side involves the same combination of variables as  the ordinary TUR does, where we make the dependence on $\dfrequ$ explicit. The right-hand side additionally contains the derivative of
the current with respect to the driving frequency, i.e., the 
response of the current to a slight change of the period of driving.
In the special case of constant driving, this bound becomes
the ordinary TUR, eq.~\eqref{eq:intro:TUR}, since then the current is formally independent of the driving frequency. Thus, the entropy production in a 
periodically driven system can be bounded using  experimental data for the mean of any current, its fluctuations and its
response to a slight change of driving frequency.

One consequence of this relation is
 that it provides  a necessary condition for (almost)
dissipation-less precision. The current must  
be proportional to the frequency of driving since then the
right-hand side vanishes. This insight rationalizes the earlier 
construction of a  dissipation-less Brownian clock~\cite{bara16}.

Applied to the efficiency $\eta\equiv P\out/P\inn$ of isothermal
cyclic engines that convert an input source of
energy with mean $P\inn$ to an output power with mean $P\out=P\inn-\sigma$, the
relation implies
\beq
P\out\leq \frac{1-\eta}{\eta}\frac {D_{{P\out}}}{[1-\dfrequ P'\out(\dfrequ)/P\out(\dfrequ)]^2} .
\ee
Hence, in general, the power of a cyclic engine vanishes  at least linearly as its efficiency approaches the maximum value of 1. A finite power in this limit is, in principle, possible only if the current fluctuations diverge or if the output power is
proportional to the cycling frequency of the engine. While the
first option has been previously been derived for steady-state
engines from the TUR~\cite{piet16b}, the second one is genuine for periodically driven engines.

These results and the further ones derived and discussed below hold for systems described by a Markov dynamics on a set of states. 
The transition rate $k_{ij}(\tau)$ from state $i$ to state $j$ is time-periodic with $k_{ij}(\tau+\mathcal{T})=k_{ij}(\tau)$ and $0\le\tau\le\period$.
In the long-time limit, the probability to find the system in state $i$ becomes periodic as well and will be denoted by $\pst{i}{\dfrequ}{\tau}$, where the corresponding driving frequency $\dfrequ$ is explicitly introduced through the second argument.

One class of fluctuating time-integrated current depends on the number of transitions between states
\beq
J_T^d\equiv \frac{1}{T}\int_0^T\dd{t}\sum_{ij} \dmij(t)d_{ij}(t)
\label{eq:intro:fluct_current}
\ee
with periodic increments $d_{ij}(t)=-d_{ji}(t)$ and $0\le t\le T$.
Here, $\mij(t)$ is the
number of transitions from $i$ to $j$ up to time $t$ along a trajectory $i(t)$ of length $T$.
In the long-time limit $T\to\infty$, the current in eq.~\eqref{eq:intro:fluct_current} reaches the mean value
\begin{equation}
J^d(\dfrequ)\equiv\expval{J_T^d}=\frac{1}{\period}\int_0^\period\dd{\tau}\sum_{i>j}\jst{\dfrequ}{\tau} d_{ij}(\tau),
\end{equation}
where $\jst{\dfrequ}{\tau}\equiv\pst{i}{\dfrequ}{\tau}k_{ij}(\tau)-\pst{j}{\dfrequ}{\tau}k_{ji}(\tau)$.
One example for the mean value of a current in eq.~\eqref{eq:intro:fluct_current} is the entropy production
\begin{equation}
\sigma(\dfrequ) = \frac{1}{\period}\int_0^\period\dd{\tau}\sum_{i>j}\jst{\dfrequ}{\tau}\ln\left(\frac{\pst{i}{\dfrequ}{\tau}k_{ij}(\tau)}{\pst{j}{\dfrequ}{\tau}k_{ji}(\tau)}\right).
\label{eq:intro:entropy_production}
\end{equation}
A further class of current can be derived from the residence time in certain states as
\begin{equation}
J_T^a \equiv \frac{1}{T}\int_0^T\dd{t}\sum_i \occi\dai,
\label{eq:intro:occ_current}
\end{equation}
where $\occi$ is 1 if state $i$ is occupied at time $t$ and 0, otherwise.
Here, the periodic increment can be written as a time-derivative of a state variable, e.g., for power input $\dai=\dot{E}_i(t)$, where $E_i(t)$ is the energy of state $i$.
The mean value of eq.~\eqref{eq:intro:occ_current} is denoted by $J^a(\dfrequ)\equiv\expval{J_T^a}$.
An arbitrary average current consists of a superposition of the two types of currents defined in eqs.~\eqref{eq:intro:fluct_current} and~\eqref{eq:intro:occ_current}, i.e., $J(\dfrequ)\equiv J^d(\dfrequ) + J^a(\dfrequ)$.

A second class of observables are called residence quantities that are defined as
\begin{equation}
A_T \equiv \frac{1}{T}\int_0^T\dd{t}\sum_i \occi\sv
\label{eq:intro:A}
\end{equation}
with time-dependent periodic state variables $A_i(t)$ that cannot be written as time-derivatives.
Simple examples of such a quantity are the average fraction of time $\tau_k$ spent in state $k$ during one cycle with increment $\sv=\delta_{i,k}$ or the average energy with $\sv=E_i(t)$.
For long times, their mean value is given by
\begin{equation}
A(\dfrequ)\equiv\expval{A_T}=\frac{1}{\period}\int_0^\period\dd{\tau}\sum_i\pst{i}{\dfrequ}{\tau}A_i(\tau).
\end{equation}
Fluctuations of residence and current observables can be quantified via the diffusion coefficient
\begin{equation}
D_X(\dfrequ) \equiv \lim\limits_{T\to\infty}T\expval{\left(X_T-\expval{X_T}\right)^2}/2
\end{equation}
with $X_T\in\{J^d_T,J^a_T,A_T\}$.

For the residence variable $A_T$ in eq.~\eqref{eq:intro:A}, we will show that the mean value $A(\dfrequ)$ and the fluctuations $D_A$ obey
\beq
 \sigma(\dfrequ)D_A(\dfrequ)\geq [\dfrequ A'(\dfrequ)]^2 .
 \label{eq:main_result_A}
 \ee 
Thus, by measuring how the mean value of this observable changes with driving frequency, a lower bound on the entropy production can be obtained without knowing further details of the system. 
It is quite remarkable that by observing a variable that is even under time-reversal a bound on entropy production, which is a hallmark of broken time-reversal symmetry, can be inferred.
There is no analog of this relation in the case of constant driving, since then the right-hand side vanishes.

As a first example, we illustrate relation~\eqref{eq:main_result_A} via an isothermal three state model with the following energy levels
\begin{align}
E_1(t) &= E^c_1\cos\dfrequ t + E^s_1\sin\dfrequ t+ E_1^0,\nonumber\\ 
E_2(t) &=  E_2^0, \quad \text{and} \quad E_3(t) = E_3^0.
\label{eq:examples:E_three_state}
\end{align}
The rates $k_{ij}(\tau)$ 
\begin{align}
k_{ij}(t) &= k^0_{ij}\exp\left[-\alpha(E_i(t)-E_j(t))\right],\\
k_{ji}(t) &= k^0_{ij}\exp\left[(1-\alpha)(E_i(t)-E_j(t))\right]
\end{align} 
fulfill the local detailed balance condition for any $\alpha$.
Here, we have set $\beta=1$ and introduced a rate amplitude $k^0_{ij}$ for each link $(i,j)$.

\begin{figure}[tbp]
\includegraphics[width=0.5\textwidth]{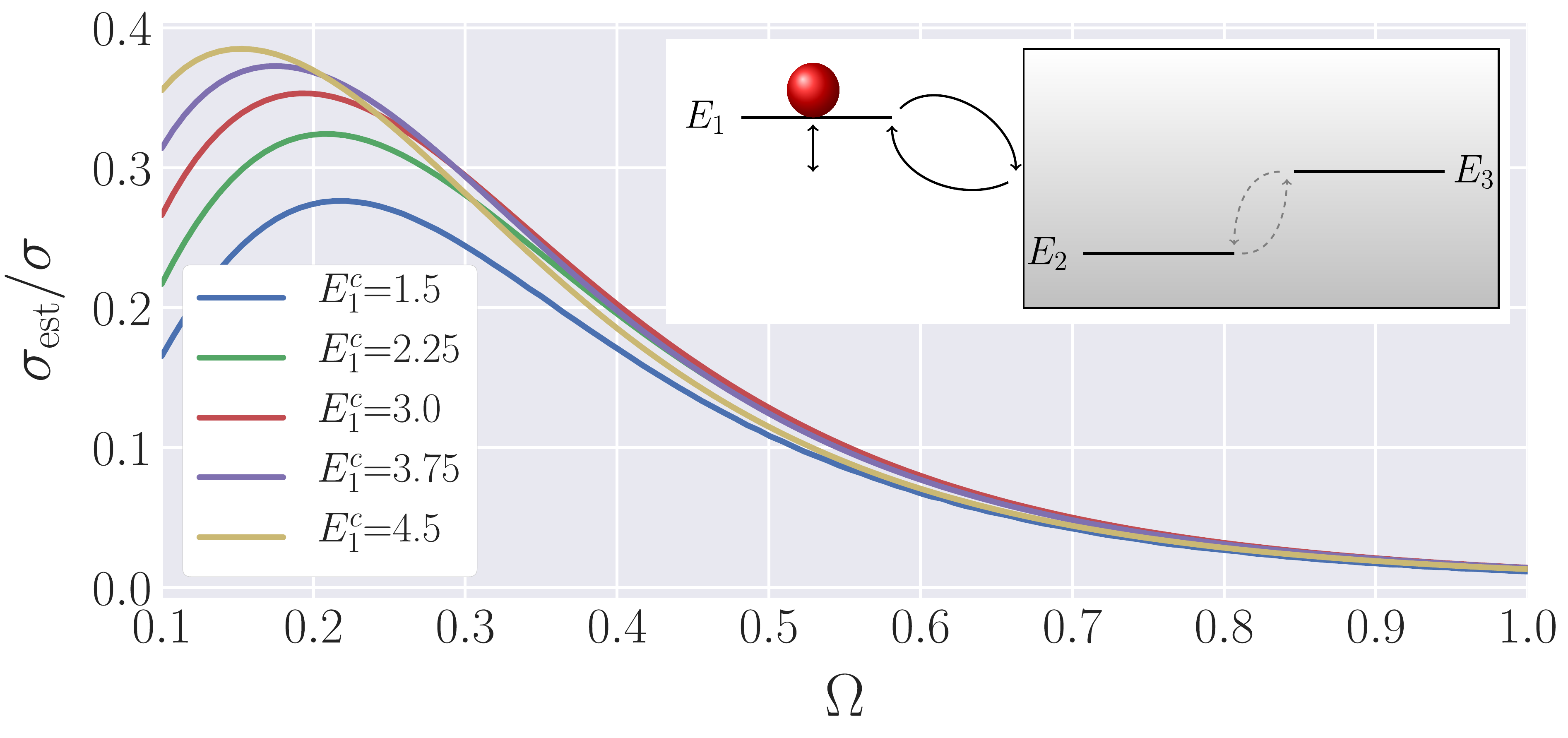}
\caption{Ratio of estimator~\eqref{eq:examples:cgm_sigma_est} and total entropy production~\eqref{eq:intro:entropy_production} as a function of the driving frequency $\dfrequ$ for different energy amplitudes $E^c_1$ for the driven 3-state level (inset).
The parameters $E^s_1=1.0,E_1^0=8.0, E_2^0=2.3, E_3^0=4.1, k^0_{12}=0.01, k^0_{23}=0.2,k^0_{13}=10.0,$ and $\alpha=0.1$ are kept fixed.}
\label{fig:examples:sojourn}
\end{figure}

We vary the driving frequency $\dfrequ$ for different energy amplitudes $E^c_1$ and fix all other parameters.
The total entropy production can be estimated via eq.~\eqref{eq:main_result_A}:
the frequency-dependent fraction of residence time $\tau_1(\dfrequ)$ in state 1 and its diffusion coefficient $D_{\tau_1}(\dfrequ)$ are sufficient to obtain the estimator
\begin{equation}
\sigma_\mathrm{est} \equiv \left[\dfrequ\tau_1'(\dfrequ)\right]^2/D_{\tau_1}(\dfrequ)\le \sigma,
\label{eq:examples:cgm_sigma_est}
\end{equation} plotted  in Fig.~\ref{fig:examples:sojourn}a as a function of the driving frequency $\dfrequ$ for different amplitudes $E^c_1$.
Hence, this simple estimate yields already up to 40\% of the total entropy production.

The derivation of our main results~\eqref{eq:main_result_J} and~\eqref{eq:main_result_A}
can be sketched as follows~\cite{notekoyu19a}.
The diffusion coefficient $D_G(\dfrequ)$ for a quantity
\begin{equation}
G_T \equiv A_T + J^a_T + J^d_T
\label{eq:main:derivation:G_def}
\end{equation}
with mean $G(\dfrequ)\equiv\expval{G_T}$ can be bounded by
\begin{equation}
2D_G(\dfrequ)\ge (G(\dfrequ)-G(\dfrequ, \dfrequaux))^2/[(\dfrequ/\dfrequaux-1)^2\gbound(\dfrequ, \dfrequaux)],
\label{eq:main:derivation:bound_omega}
\end{equation}
where $\dfrequaux$ is an arbitrary frequency and
\begin{align}
\gbound(\dfrequ, \dfrequaux) &\equiv \frac{1}{\period}\int_0^\period\dd{\tau}\sum_{i>j}\left[\jst{\dfrequaux}{\dfrequ\tau/\dfrequaux}\right]^2/\tst{\dfrequ}{\tau},\\
G(\dfrequ, \dfrequaux) &\equiv  A(\dfrequaux)+(\dfrequ/\dfrequaux) J(\dfrequaux)
\end{align}
with $J(\dfrequaux)\equiv J^a(\dfrequaux) + J^d(\dfrequaux)$.
Here, we introduced the probability current at frequency $\dfrequaux$
\begin{equation}
\jst{\dfrequaux}{\dfrequ\tau/\dfrequaux}\equiv \pst{i}{\dfrequaux}{\dfrequ\tau/\dfrequaux}k_{ij}(\tau)-\pst{j}{\dfrequaux}{\dfrequ\tau/\dfrequaux}k_{ji}(\tau)
\end{equation}
and the activity at link $(i,j)$
\begin{equation}
\tst{\dfrequ}{\tau} \equiv \pst{i}{\dfrequ}{\tau}k_{ij}(\tau)+\pst{j}{\dfrequ}{\tau}k_{ji}(\tau)
\end{equation} 
at the original frequency $\dfrequ$.

We choose either the increments $d_{ij} =\dot{a}_i=0$ or $A_i=0$ such that $G(\dfrequ)=J(\dfrequ)$ for currents and $G(\dfrequ)=A(\dfrequ)$ for residence quantities and take the limit $\dfrequaux\to\dfrequ$.
This leads to the following two inequalities 
\begin{align}
D_J(\dfrequ)\bound(\dfrequ)/J(\dfrequ)^2 &\ge (1-\dfrequ J'(\dfrequ)/J(\dfrequ))^2,\label{eq:main:derivaton:main_res_J}\\
D_A(\dfrequ)\bound(\dfrequ)/A(\dfrequ)^2 &\ge (\dfrequ A'(\dfrequ)/A(\dfrequ))^2.\label{eq:main:derivaton:main_res_A}
\end{align}
Here, $\bound(\dfrequ)\in\set{2\gbound(\dfrequ), 2\act(\dfrequ), \sigma(\dfrequ)}$ can be chosen as one of three cost terms with $\gbound(\dfrequ)\equiv\gbound(\dfrequ,\dfrequaux=\dfrequ)$ and $\act(\dfrequ)\equiv 1/\period\int_0^\period\dd{\tau}\sum_{i>j}\tst{\dfrequ}{\tau}$ the average dynamical activity.
These two inequalities~\eqref{eq:main:derivaton:main_res_J} and~\eqref{eq:main:derivaton:main_res_A} are our most general results, from which our main results~\eqref{eq:main_result_J} and~\eqref{eq:main_result_A} are obtained with $\bound(\dfrequ)=\sigma(\dfrequ)$.
 
By choosing $\bound(\dfrequ)=\act(\dfrequ)$, the average dynamical activity can be bounded via eq.~\eqref{eq:main:derivaton:main_res_J} by current observables and their fluctuations.
The steady-state analog of this relation was proven in~\cite{piet15} and extended to more general observables in~\cite{garr17}.
Our generalization of the latter one for periodic driving is derived in the Supplemental Material~\cite{notekoyu19a}.
These bounds on activity can be generalized to residence observables through eq.~\eqref{eq:main:derivaton:main_res_A}.
They have no steady-state analog.

Finally, we investigate the implications of our results for heat engines operating cyclically between two baths of inverse temperature $\beta_h<\beta_c$ with efficiency
\begin{equation}
\eta(\dfrequ)\equiv P_\mathrm{out}(\dfrequ)/\dot{Q}_h(\dfrequ)\le \eta_C \equiv 1 - \beta_h/\beta_c,
\end{equation}
where $P_\mathrm{out}>0$ is the output power and $\dot{Q}_h$ the heat current flowing into the system from the hot reservoir. The inequality~\eqref{eq:main_result_J} implies the bound 
\begin{equation}
\eta(\dfrequ) \le \hat{\eta}(\dfrequ) \equiv \eta_C\left(1+\frac{(P_\mathrm{out}(\dfrequ)-\dfrequ P'_\mathrm{out}(\dfrequ))^2}{(\beta_c D_{P_\mathrm{out}(\dfrequ)}P_\mathrm{out}(\dfrequ))}\right)^{-1}.
\label{eq:examples:eta_bound}
\end{equation} 
Carnot efficiency at finite power
can thus be reached only if the power fluctuations diverge or if the power increases
linearly with the driving frequency. The latter condition is typical for quasi-static driving~\cite{holu18}.
For maximal output power ($P'_\mathrm{out}(\dfrequ_\mathrm{max})=0$), eq.~\eqref{eq:examples:eta_bound} reduces to the established bound for steady-state systems~\cite{piet17a} thus showing
a universal trade-off between efficiency, power and constancy at maximum power.

So far we discussed systems with a discrete set of states. 
We now illustrate the new bound on efficiency~\eqref{eq:examples:eta_bound} for a system obeying an overdamped Langevin equation.
Specifically, we consider a Stirling heat engine model inspired by~\cite{blic12,schm08}.
The engine consists of a colloid in a one-dimensional harmonic potential 
with a time-dependent stiffness $\lambda(\tau)$
\begin{equation}
V(x, \lambda(\tau)) \equiv V(x, \tau) \equiv \lambda(\tau)x^2/2\ge0.
\label{eq:examples:V_x}
\end{equation}
The position $x$ follows the
 Langevin equation
\begin{equation}
\dot{x}(t) = -\mu \lambda(t) x(t) + \zeta(t).
\label{eq:examples:langevin}
\end{equation}
Here, $\mu$ denotes the mobility, and $\zeta(t)$ is zero-mean Gaussian noise with correlation $\expval{\zeta(t)\zeta(\tilde{t})}=2\mu\delta(t-\tilde{t})/\beta(t)$ and a periodic temperature $\beta(t)=\beta(t+\period)$. 
The corresponding Fokker-Planck equation for the probability distribution $p(x,t)$ reads
\begin{equation}
\partial_\tau p(x,t) = \mu\partial_x \left(\partial_x V(x,t)+\beta^{-1}(t)\partial_x\right)p(x,t).
\label{eq:examples:current}
\end{equation}
The periodic stationary solution $p^\mathrm{ps}(x,\tau)$ of \eqref{eq:examples:current} is a Gaussian with zero mean.

\begin{figure}[tbp]
\centering\includegraphics[width=0.5\textwidth]{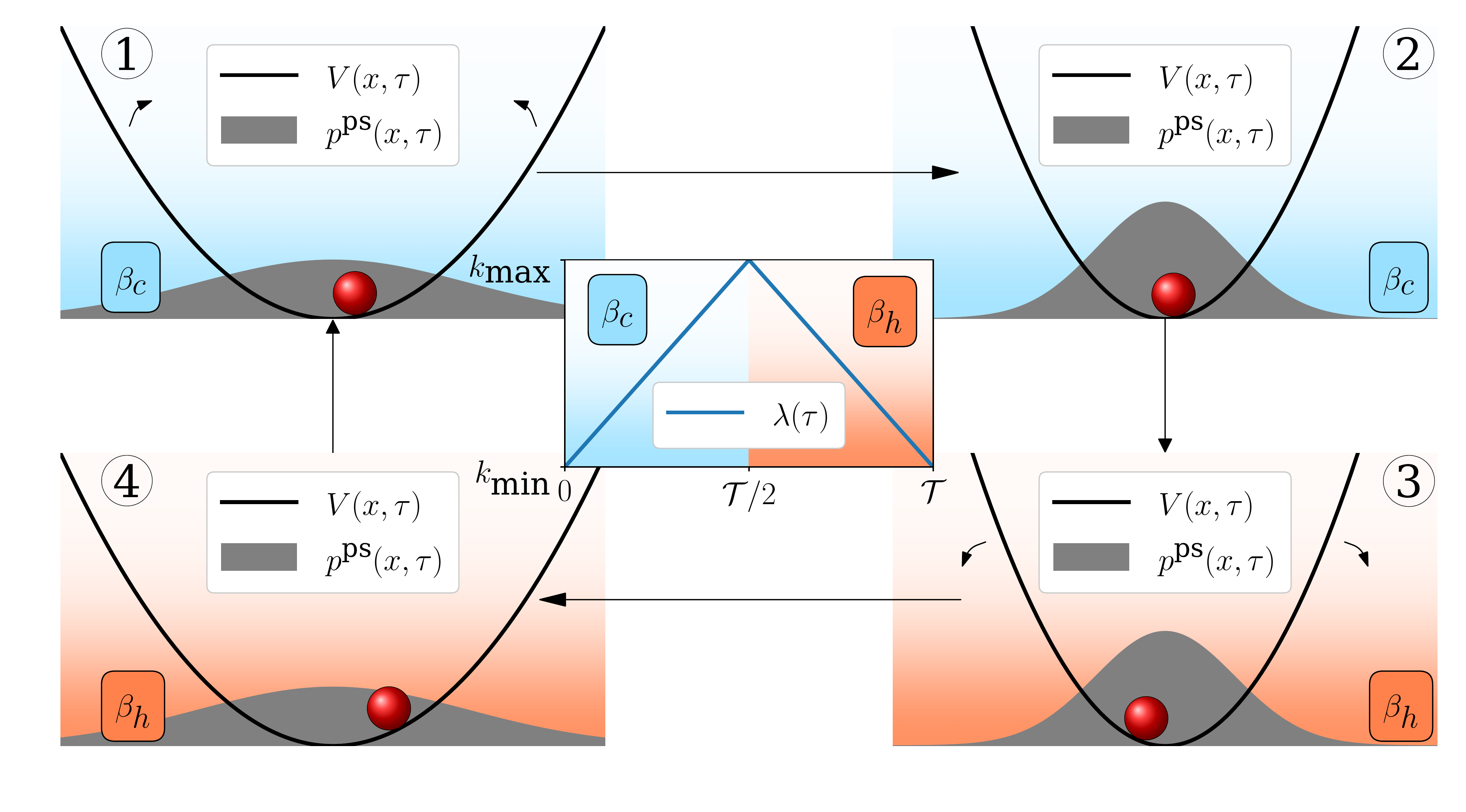}
\caption{Particle in a harmonic trap with time-dependent stiffness $\lambda(\tau)$ operating as a Stirling heat engine between temperatures $\beta_c>\beta_h$. 
}
\label{fig:exmaples:stirling_engine}
\end{figure}

As a protocol for the stiffness, we use the one from the experiment in~\cite{blic12}, which  increases and decreases linearly in time according to
\begin{equation}
\lambda(\tau) = 
\begin{cases}
2\Delta k\tau/\period+k_\mathrm{min}, &0 \le \tau < \period/2\\
-2\Delta k(\tau/\period-1/2)+k_\mathrm{max},& \period/2 \le\tau <\period
\end{cases}
\label{eq:examples:protocol}
\end{equation}
with $\Delta k \equiv k_\mathrm{max}-k_\mathrm{min}$, see Fig.~\ref{fig:exmaples:stirling_engine}.
The coupling to two different heat baths leads to a time-dependent temperature
that is piecewise constant, i.e., $\beta(\tau)=\beta_c$ for $0 \le \tau < \period/2$ and $\beta(\tau)=\beta_h$ for $\period/2\le\tau <\period$.

\begin{figure}[tbp]
\includegraphics[width=0.5\textwidth]{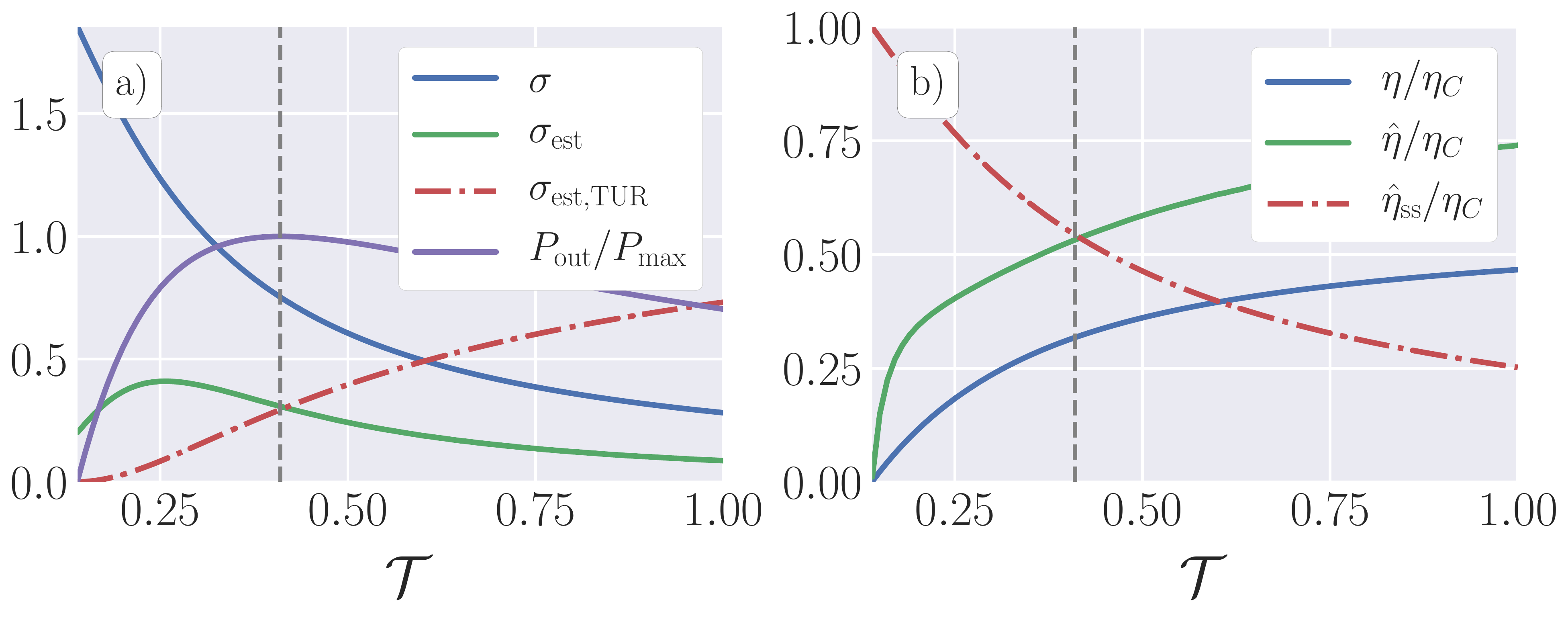}
\caption{(a) Entropy production $\sigma$, its corresponding estimator $\sigma_\mathrm{est}$~\eqref{eq:examples:he_sigma_est}, estimator for steady-state systems $\sigma_\mathrm{est,TUR}$ and output power $P\out$ divided by its maximum $P_\mathrm{max}$ and (b) efficiency $\eta$, bound on the efficiency $\hat{\eta}$~\eqref{eq:examples:eta_bound}, and its steady-state analog $\hat{\eta}_\mathrm{ss}$ as function of the cycle duration $\period$ for $\mu=10.0$, $k_\mathrm{max}=2.0$, $k_\mathrm{min}=1.0$, $\beta_c=2.0$ and $\beta_h=1.0$.}
\label{fig:examples:stirling_he}
\end{figure}

Currents of interest are power, entropy production and heat.
Their explicit expressions and the diffusion coefficient for the output power are given in \cite{notekoyu19a}.
In order to illustrate eq.~\eqref{eq:main_result_J} and~\eqref{eq:examples:eta_bound}, we define the estimator
\begin{equation}
\sigma_\mathrm{est} \equiv \left(P_\mathrm{out}(\dfrequ)-\dfrequ P'_\mathrm{out}(\dfrequ)\right)^2/D_{P\out}(\dfrequ)\le \sigma
\label{eq:examples:he_sigma_est}
\end{equation}
for entropy production and vary the cycle duration $\period$. 
The output power, the estimator~\eqref{eq:examples:he_sigma_est} and the bound on efficiency~\eqref{eq:examples:eta_bound} are shown in Fig.~\ref{fig:examples:stirling_he}.
With increasing $\period$, the output power increases for small $\period$.
According to eq.~\eqref{eq:main_result_J} the ordinary TUR, eq.~\eqref{eq:intro:TUR}, is still valid.
At maximum power (vertical dotted line) eq.~\eqref{eq:main_result_J} is equivalent to the TUR and the estimator $\sigma_\mathrm{est,TUR}\equiv P\out^2/D_{P\out}$ intersects $\sigma_\mathrm{est}$.
After reaching the maximum value, the power decreases with increasing $\period$ with the TUR violated for $\period\gtrsim 0.6$.

In summary, for periodically driven systems,  we have derived a class of inequalities that relate, \textsl{inter alia},  entropy production with the mean of an observable, its dependence on driving frequency and its variance. Remarkably, apart from the more familiar currents,
such observables can also be even under time-reversal like residence times are. For
cyclic heat engines and cyclically driven isothermal engines, the thermodynamic efficiency can now be bounded using data only for the input or output current. Since our results have been obtained for a Markov dynamics on a discrete set of states, they trivially
hold as well for overdamped Langevin dynamics. It remains an open problem whether they can be extended
to underdamped dynamics. Arguably even more exciting will be to explore
along similar lines  periodically driven open quantum systems for which coherences are relevant.

\bibliography{./refs}
\end{document}


\preprint{APS/123-QED}
\title{Supplemental Material for "Operationally accessible bounds on fluctuations
	and entropy production
	in periodically driven systems"\\~\\}

\author{Timur Koyuk and Udo Seifert\\~
}
\affiliation{
{II.} Institut f\"ur Theoretische Physik, Universit\"at Stuttgart,
  70550 Stuttgart, Germany
}
\date{\today}

\maketitle
\section{I. Derivation of the main results\label{sec:derivation}}
We derive our main results by using a lower bound on the scaled cumulant generating function (SCGF)~\cite{dech17,koyu19}
\begin{equation}
\lambda_T(z)\equiv\frac{1}{T}\ln\expval{e^{zTX_T}}.
\label{eq:appendix:derivation:scgf_def}
\end{equation}
Here, $X_T$ is an arbitrary observable defined along a trajectory $i(\tau)$ of length $T$.
The bound on the generating function is expressed in terms of an auxiliary dynamics.
This dynamics is described by rates  $\mathbf{\tilde{k}}(\tau)\equiv \set{\tilde{k}_{ij}(\tau)}$ leading to densities $\vbpaux \equiv \set{\paux}$.
The average with respect to the auxiliary dynamics is denoted by $\expvalaux{\cdot}$.
A general bound on the SCGF at time $T=n\period$ (see, e.g.,~\cite{koyu19}) is given by
\begin{widetext}
\begin{equation}
\lambda_{n\period}(z) \ge z\expvalaux{X_T} - F[\vbpaux, \mathbf{\tilde{k}}(\tau)] - \frac{1}{n\period} D\left(\mathbf{\tilde{p}}(0)\vert\vert \mathbf{p^\mathrm{ps}}(0;\dfrequ)\right),
\label{eq:derivation:scgf_bound_k}
\end{equation}
with
\begin{align}
F[\vbpaux, \mathbf{\tilde{k}}(\tau)]\equiv \frac{1}{\period}\int_0^\period\dd{\tau}\sum_{i,j}\Bigg(&\paux\tilde{k}_{ij}(\tau)\ln\left(\frac{\tilde{k}_{ij}(\tau)}{k_{ij}(\tau)}\right)
-\paux[\tilde{k}_{ij}(\tau)-k_{ij}(\tau)]\Bigg)
\label{eq:derivaton:F}
\end{align}
and the  Kullback-Leibler divergence
\begin{equation}
D\left(\mathbf{\tilde{p}}(0)\vert\vert \mathbf{p^\mathrm{ps}}(0;\dfrequ)\right)\equiv\sum_i\tilde{p}_i(0)\ln[\tilde{p}_i(0)/p^\mathrm{ps}_i(0;\dfrequ)].
\end{equation}
\end{widetext}
Here, $k_{ij}(\tau)$ are rates of the original dynamics with frequency $\dfrequ$ whose periodic stationary state is denoted by $\pst{i}{\dfrequ}{\tau}$.

After an optimization with respect to the activity of the auxiliary dynamics (see, e.g.,~\cite{koyu19}), this bound is expressed in terms of  densities $\vbpaux \equiv \set{\paux}$ and currents $\vbjaux \equiv \set{\jaux}$ by choosing optimal rates $\tilde{k}^*_{ij}(\tau)$.
For choosing $X_T=G_T$ the bound is given by
\begin{widetext}
\begin{equation}
\lambda_{n\period}(z) \ge z\expvalaux{G_T} - \frac{1}{\period}\int_0^\period\dd{\tau}L\left(\vbpaux, \vbjaux\right)  - \frac{1}{n\period} D\left(\mathbf{\tilde{p}}(0)\vert\vert \mathbf{p^\mathrm{ps}}(0;\dfrequ)\right),
\label{eq:derivation:scgf_bound}
\end{equation}
where the integrand $L\left(\vbpaux, \vbjaux\right)$ is defined as
\begin{align}
L\left(\vbpaux, \vbjaux\right) \equiv  &\sum_{i>j} \jaux \left[\arsinh{\frac{\jaux}{\apaux}}-\arsinh{\frac{\jpaux}{\apaux}} \right]-\left(\sqrt{[\jaux]^2+[\apaux]^2}-\sqrt{[\jpaux]^2+[\apaux]^2}\right),
\end{align}
with
\begin{align}
\jpaux \equiv \pauxi{i}k_{ij}(\tau)-\pauxi{j}k_{ji}(\tau), \quad \apaux \equiv \sqrt{4\pauxi{i}\pauxi{j}k_{ij}(\tau)k_{ji}(\tau)}.
\end{align}
\end{widetext}

The above introduced auxiliary dynamics has to describe a physical process and hence, it has to obey the following conditions
\begin{align}
\sum_i\paux &= 1,& &0<\paux<1,\label{eq:derivation:aux_dyn_conditions_1}\\
\pauxt{0}&=\pauxt{\period},& &\dpaux = -\sum_j\jaux,
\label{eq:derivation:aux_dyn_conditions}
\end{align}
with probability current
\begin{equation}
\jaux\equiv \tilde{p}_i(\tau) \tilde{k}_{ij}(\tau)-\tilde{p}_j(\tau) \tilde{k}_{ji}(\tau)
\end{equation}
in the auxiliary dynamics.
The first condition in~\eqref{eq:derivation:aux_dyn_conditions} guarantees that the chosen density $\paux$ has the same cycle duration $\period$ as the original periodic stationary state.
The latter one can be written as
\begin{align}
\pst{i}{\dfrequ}{\tau} \equiv a_0(\dfrequ) + \sum_{n=1} \Big(&a_n(\dfrequ) \cos(n\dfrequ \tau)\nonumber\\ 
&+ b_n(\dfrequ)\sin(n\dfrequ \tau) \Big).
\label{eq:derivation:ps_fourier_series}
\end{align}
For every frequency $\dfrequ$ there exists a corresponding set of Fourier coefficients $\set{a_n(\dfrequ), b_n(\dfrequ)}$ defining the unique periodic stationary state $\pst{i}{\dfrequ}{\tau}$.

We choose the following ansatz for the auxiliary dynamics
\begin{align}
\paux &= \pst{i}{\dfrequ}{\tau} + \epsilon \left(\pst{i}{\dfrequ}{\tau}  - \pst{i}{\dfrequaux}{\dfrequ\tau/\dfrequaux}\right),\nonumber\\ 
\jaux &= \jst{\dfrequ}{\tau} + \epsilon \left(\jst{\dfrequ}{\tau} - \dfrequ/\dfrequaux\jst{\dfrequaux}{\dfrequ\tau/\dfrequaux}\right)
\label{eq:derivation:ansatz}
\end{align}
with small $\epsilon=\order{z}$, where the density
\begin{align}
\pst{i}{\dfrequaux}{\dfrequ\tau/\dfrequaux} \equiv a_0(\dfrequaux) + \sum_{n=1} \Big(&a_n(\dfrequaux) \cos(n\dfrequ \tau)\nonumber\\
 &+ b_n(\dfrequaux)\sin(n\dfrequ \tau) \Big)
\end{align}
has the same cycle duration $\period=2\pi/\dfrequ$ as the original process.
The Fourier coefficients $\set{a_n(\dfrequaux), b_n(\dfrequaux)}$ correspond to a periodic stationary process with frequency $\dfrequaux$. 

Inserting ansatz \eqref{eq:derivation:ansatz} into \eqref{eq:derivation:scgf_bound}, taking the long time limit $n\to\infty$, choosing small $\epsilon=\order{z}$, and optimizing  with respect to this parameter leads to a local quadratic bound on the generating function, which implies the inequality
\begin{equation}
2D_G(\dfrequ)\ge \frac{(G(\dfrequ)-G(\dfrequ, \dfrequaux))^2}{(\dfrequ/\dfrequaux-1)^2\gbound(\dfrequ, \dfrequaux)}
\label{eq:derivation:bound_omega}
\end{equation}
with
\begin{align*}
\gbound(\dfrequ, \dfrequaux) &\equiv \frac{1}{\period}\int_0^\period\dd{\tau}\left(\sum_{i>j}\frac{\left(\jst{\dfrequaux}{\dfrequ\tau/\dfrequaux}\right)^2}{t^\mathrm{ps}_{ij}(\tau;\dfrequ)}\right),\\
G(\dfrequ, \dfrequaux) &\equiv  \dfrequ/\dfrequaux J(\dfrequaux) + A(\dfrequaux),\nonumber\\
G(\dfrequ) &\equiv G(\dfrequ, \dfrequaux=\dfrequ)=J(\dfrequ) + A(\dfrequ)
\end{align*}
as quoted in the main text.
The nominator in~\eqref{eq:derivation:bound_omega} follows from the first term in~\eqref{eq:derivation:scgf_bound}, i.e., the average of observable $G_T$ in the auxiliary dynamics.
Here, we used the substitution $\tilde{\tau}=\dfrequ\tau/\dfrequaux$ and the following properties of the increments
\begin{align}
\dij{ij}{\dfrequ}{\tau} &= \dij{ij}{\dfrequ}{\dfrequaux\tilde{\tau}/\dfrequ}=\dij{ij}{\dfrequaux}{\tilde{\tau}},\nonumber\\
\ai{i}{\dfrequ}{\tau} &= (\dfrequ/\dfrequaux)\ai{i}{\dfrequ}{\dfrequaux\tilde{\tau}/\dfrequ}=(\dfrequ/\dfrequaux)\ai{i}{\dfrequaux}{\tilde{\tau}}\nonumber,\\
\Ai{i}{\dfrequ}{\tau} &= \Ai{i}{\dfrequ}{\dfrequaux\tilde{\tau}/\dfrequ}=\Ai{i}{\dfrequaux}{\tilde{\tau}},\nonumber
\end{align}
where we introduced the second argument to denote the corresponding frequency.
The denominator in~\eqref{eq:derivation:bound_omega}, i.e., the cost term, follows from the second term in eq.~\eqref{eq:derivation:scgf_bound}.
After taking the long-time limit, the third term in eq.~\eqref{eq:derivation:scgf_bound}, i.e., the Kullback-Leibler divergence, vanishes.
Moreover, we note that for $\dfrequaux\to\infty$ and $\Ai{i}{\dfrequ}{\tau}=0$ in eq.~\eqref{eq:derivation:bound_omega} the GTUR in \cite{koyu19} is recovered.

Next, we choose $\dfrequaux=\alpha\dfrequ$ and take the limit $\alpha\to 1$.
Both, nominator and denominator on the right-hand side in eq.~\eqref{eq:derivation:bound_omega} approach zero in this limit.
Hence, we expand both around $\alpha=1$ leading to
\begin{equation}
D_G(\dfrequ)\ge \frac{x^2\left(J(\dfrequ)-\dfrequ J'(\dfrequ) - \dfrequ A'(\dfrequ)+\order{x}\right)^2}{\left(x^2+\order{x^3}\right) \left(\gbound(\dfrequ)+\order{x}\right)}
\end{equation}
with $x\equiv\alpha-1$ and
\begin{equation}
\gbound(\dfrequ) \equiv \gbound(\dfrequ,\dfrequaux=\dfrequ)= \frac{1}{\period}\int_0^\period\dd{\tau}\left(\sum_{i>j}\frac{\left(\jst{\dfrequ}{\tau}\right)^2}{t^\mathrm{ps}_{ij}(\tau;\dfrequ)}\right).
\end{equation}
Here, $'$ denotes the derivative with respect to frequency $\dfrequ$.
Moreover, by setting either $\dij{ij}{\dfrequ}{\tau}=\ai{i}{\dfrequ}{\tau}=0$ or $\Ai{i}{\dfrequ}{\tau}=0$ we get the following inequalities for currents $J(\dfrequ)$ and residence quantities $A(\dfrequ)$
\begin{align}
2D_J(\dfrequ)\gbound(\dfrequ)/J(\dfrequ)^2 &\ge (1-\dfrequ J'(\dfrequ)/J(\dfrequ))^2,\\
2D_A(\dfrequ)\gbound(\dfrequ)/A(\dfrequ)^2 &\ge (\dfrequ A'(\dfrequ)/A(\dfrequ))^2,
\end{align}
as quoted in the main text.

Finally, using $2\jps^2\le \tps \sigma_{ij}(\tau)$ and $\jps^2\le \tps^2$ implies our main results
\begin{align}
D_J(\dfrequ)\bound(\dfrequ)/J(\dfrequ)^2 &\ge (1-\dfrequ J'(\dfrequ)/J(\dfrequ))^2,\label{eq:derivaton:main_res_J}\\
D_A(\dfrequ)\bound(\dfrequ)/A(\dfrequ)^2 &\ge (\dfrequ A'(\dfrequ)/A(\dfrequ))^2,\label{eq:derivaton:main_res_A}
\end{align}
where $\bound(\dfrequ)\in\set{2\gbound(\dfrequ), 2\act(\dfrequ), \sigma(\dfrequ)}$ are cost terms and $\act(\dfrequ)\equiv 1/\period\int_0^\period\dd{\tau}\sum_{i>j}\tps$ is the average dynamical activity.

\section{II. Finite-time Generalization}
For a finite-time generalization of relation~\eqref{eq:derivation:bound_omega} and hence, of our main results~\eqref{eq:derivaton:main_res_J} and~\eqref{eq:derivaton:main_res_A}, an additional cost term arises due to the non-vanishing Kullback-Leibler divergence, i.e.,
\begin{align}
D_J(n\period;\dfrequ)[\bound(\dfrequ)+\tilde{\bound}(\dfrequ)]/J(\dfrequ)^2 &\ge (1-\dfrequ J'(\dfrequ)/J(\dfrequ))^2,\label{eq:derivaton:main_res_J_finite_time}\\
D_A(n\period;\dfrequ)[\bound(\dfrequ)+\tilde{\bound}(\dfrequ)]/A(\dfrequ)^2 &\ge (\dfrequ A'(\dfrequ)/A(\dfrequ))^2,\label{eq:derivaton:main_res_A_finite_time}
\end{align}
where the additional term
\begin{equation}
\tilde{\bound}(\dfrequ)\equiv\frac{1}{n\period}\sum_{i}\dfrequ^2[\pst{i}{\dfrequ}{0}']^2/\pst{i}{\dfrequ}{0}
\label{eq:derivation:kld_finite_time}
\end{equation}
is the contribution of the non-vanishing Kullback-Leibler divergence and
\begin{equation}
D_X(n\period;\dfrequ)\equiv n\period\expval{\left(X_{n\period}-\expval{X_{n\period}}\right)^2}/2=\lambda_{n\period}''(0)/2
\end{equation}
is the finite-time generalization of the diffusion coefficient for an observable $X_{n\period}$ after $n$ periods.
Here, $\lambda_{n\period}''(0)$ denotes the second derivative of the SCGF with respect to $z$ at $z=0$.

\section{III. Bounds on Activity}
We now generalize the bounds on activity $\act(\dfrequ)$ in the main text to arbitrary observables depending on the number of transitions 
\begin{equation}
\mathcal{X}_T \equiv \frac{1}{T}\int_0^T\dd{t}\sum_{ij}\dot{n}_{ij}(t)g_{ij}(t)\label{eq:derivation:act_X_T}
\end{equation}
along a trajectory $i(t)$ of length $T$, where $g_{ij}(t)$ are arbitrary time-periodic increments.
Note, that we do not restrict $g_{ij}(t)$ to be antisymmetric or symmetric.
For example, $\mathcal{X}_T$ can be the average number of jumps during one period from state $1\to2$ with increments $g_{ij}(t)= \delta_{1,i}\delta_{2,j}$.
The average of~\eqref{eq:derivation:act_X_T} is denoted by $\mathcal{X}(\dfrequ)\equiv\expval{\mathcal{X}_T}$.

To derive the bound on the activity, we use the lower bound~\eqref{eq:derivation:scgf_bound_k} on the generating function $\lambda_{n\period}(z)$ for the quantity $X_T=\mathcal{X}_T$.
We choose the ansatz
\begin{align}
\paux &= \pst{i}{\dfrequaux}{\dfrequ\tau/\dfrequaux},\nonumber\\ 
\tilde{k}_{ij}(\tau) &= (\dfrequ/\dfrequaux)k_{ij}(\tau) -  (\dfrequ/\dfrequaux-1)k_{ij}(\tau)\alpha_{ij}(\tau)\delta,
\label{eq:derivation:ansatz_k}
\end{align}
with
\begin{equation}
\alpha_{ij}(\tau)\equiv\sqrt{\frac{\pst{j}{\dfrequaux}{\dfrequ\tau/\dfrequaux}k_{ji}(\tau)}{\pst{i}{\dfrequaux}{\dfrequ\tau/\dfrequaux}k_{ij}(\tau)}}
\end{equation}
and $\delta$ will be chosen as 0 or 1.
For $\delta=0$, the first term in~\eqref{eq:derivation:scgf_bound_k} is given by
\begin{equation}
\expvalaux{\mathcal{X_T}}=\frac{(\dfrequ/\dfrequaux)}{\period}\int_0^\period\dd{\tau}\sum_{ij}\pst{i}{\dfrequaux}{\dfrequ\tau/\dfrequaux}k_{ij}(\tau)\hat{d}_{ij}(\tau)
\end{equation}
Using the substitution $\tilde{\tau}=\dfrequ\tau/\dfrequaux$ and the property of the increments
\begin{align}
g_{ij}(\tau;\dfrequ) = g_{ij}(\dfrequaux\tilde{\tau}/\dfrequ;\dfrequ)=g_{ij}(\tilde{\tau};\dfrequaux),
\end{align}
where the upper index denotes the corresponding frequency, leads to
\begin{equation}
\expvalaux{\mathcal{X}_T} = (\dfrequ/\dfrequaux)\mathcal{X}(\dfrequaux).
\label{eq:dervation:act_X_T_aux}
\end{equation}
Inserting~\eqref{eq:derivation:ansatz_k} into eq.~\eqref{eq:derivaton:F}, choosing $\dfrequaux=(1+\epsilon)\dfrequ$ with small parameter $\epsilon=\order{z}$ leads to
\begin{widetext}
\begin{equation}
F[\vbpaux, \mathbf{\tilde{k}}(\tau)] = \frac{1}{\period}\int_0^\period\dd{\tau}\sum_{ij}\left(\sqrt{\pst{i}{\dfrequ}{\tau}k_{ij}(\tau)}-\delta\sqrt{\pst{j}{\dfrequ}{\tau}k_{ji}(\tau)}\right)^2\frac{\epsilon^2}{2}+\order{\epsilon^3}.
\end{equation}
\end{widetext}
Choosing $\delta=0$, expanding~\eqref{eq:dervation:act_X_T_aux} for small $\epsilon$ and optimizing with respect to this parameter leads to a quadratic bound on the generating function.
Taking the long-time limit $n\to\infty$, this quadratic bound implies
\begin{equation}
2D_{\mathcal{X}}(\dfrequ)\mathcal{A}(\dfrequ)/\mathcal{X}^2(\dfrequ)\ge (1-\mathcal{X}'(\dfrequ)/\mathcal{X}(\dfrequ))^2,
\label{eq:derivation:act_bound}
\end{equation}
which is a generalization of the bound on activity in~\cite{garr17} to periodically driven systems.
We note, that for $\delta=1$, the main results for currents $J_T$ and residence quantities $A_T$ can be recovered.

Finally, the finite-time generalization of eq.~\eqref{eq:derivation:act_bound} is given by
\begin{equation}
2D_{\mathcal{X}}(n\period;\dfrequ)[\mathcal{A}(\dfrequ)+\tilde{\bound}(\dfrequ)]/\mathcal{X}^2(\dfrequ)\ge (1-\mathcal{X}'(\dfrequ)/\mathcal{X}(\dfrequ))^2,
\label{eq:derivation:act_bound_finite_time}
\end{equation}
where $\tilde{\bound}(\dfrequ)$ is defined in eq.~\eqref{eq:derivation:kld_finite_time}.

\section{IV. Stirling Heat Engine}
In this section, we calculate currents of interest and the diffusion coefficient for the output power for the Stirling engine discussed in the main text.
The periodic stationary solution of the Fokker-Planck equation in the main text is a Gaussian with zero mean
\begin{equation}
p^\mathrm{ps}(x,\tau) = \frac{1}{\sqrt{2\pi\expval{x^2(\tau)}}}\exp\left(-\frac{x^2}{2\expval{x^2(\tau)}}\right),
\label{eq:examples:ps_x}
\end{equation}
where the variance can be calculated according to
\begin{equation}
\partial_\tau\expval{x^2(\tau)}=2\mu\left(\beta^{-1}(\tau)-\lambda(\tau)\expval{x^2(\tau)} \right),
\label{eq:examples:var_dgl}
\end{equation}
which follows from the Fokker-Planck equation~\cite{schm08}.
Here, the variance has to be periodic in time, i.e. 
\begin{equation*}
\expval{x^2(0)}=\expval{x^2(\period)}.
\end{equation*}

The output power is defined by
\begin{equation}
P_\mathrm{out}\equiv -\frac{1}{\period}\int_0^\period\dd{\tau}\int_{-\infty}^\infty\dd{x}p^\mathrm{ps}(x,\tau)\dot{V}(x,\tau).
\label{eq:examples:power}
\end{equation}
The entropy production associated  with heat dissipation in the medium given by
\begin{equation}
\sigma_m \equiv \frac{1}{\period}\int_0^\period\dd{\tau}\int_{-\infty}^\infty\dd{x} \beta(\tau)j^\mathrm{ps}(x,\tau) F(x,\tau),
\label{eq:examples:entropy_production}
\end{equation}
where
\begin{equation}
j^\mathrm{ps}(x,\tau)\equiv -\mu\left[\partial_xV(x,\tau)+\beta^{-1}(\tau)\partial_x\right]p^\mathrm{ps}(x,\tau)
\end{equation}
is the probability current.
If the system has reached the periodic stationary state, the entropy production $\sigma_m$ of the medium averaged over one cycle duration $\period$ is identical to the average total entropy production of the system.
Hence, we use the notation $\sigma\equiv\sigma_m$.
The heat current flowing into or out of the system from or into the heat bath is defined by
\begin{equation}
\dot{Q} \equiv \frac{1}{\period}\int\dd{\tau}\int_{-\infty}^\infty\dd{x} j^\mathrm{ps}(x,\tau) F(x,\tau).
\label{eq:examples:heat_current}
\end{equation}
Note that the integration over $\tau$ is performed during the time interval at which the heat bath is connected to the system.

Inserting \eqref{eq:examples:ps_x} into \eqref{eq:examples:power},~\eqref{eq:examples:entropy_production} and \eqref{eq:examples:heat_current} yields the following expressions for power
\begin{equation}
P_\mathrm{out} = -\frac{1}{\period}\int_0^\period\dd{\tau}\frac{\dot{\lambda}(\tau)}{2}\expval{x^2(\tau)},
\label{eq:examples:P_out}
\end{equation}
entropy production
\begin{equation*}
\sigma = \frac{1}{\period}\int_0^\period\dd{\tau}\frac{\mu\lambda(\tau)}{2}\left(\beta(\tau)\lambda(\tau)\expval{x^2(\tau)}-1\right),
\end{equation*}
and the cold heat flux flowing out of the system
\begin{align}
\dot{Q}_c &= \frac{1}{\period}\int_0^{\period/2}\dd{\tau}\frac{\mu\lambda(\tau)}{2\beta_c}\left(\beta_c\lambda(\tau)\expval{x^2(\tau)}-1\right),
\end{align}
as well as the hot heat flux flowing into the system
\begin{align}
\dot{Q}_h &= -\frac{1}{\period}\int_{\period/2}^{\period}\dd{\tau}\frac{\mu\lambda(\tau)}{2\beta_h}\left(\beta_h\lambda(\tau)\expval{x^2(\tau)}-1\right).
\label{eq:examples:Q_h}
\end{align}
These currents of interest can be calculated by solving~\eqref{eq:examples:var_dgl} for the initial condition leading to the periodic stationary state.

The diffusion coefficient of the output power~\eqref{eq:examples:P_out} is given by
\begin{equation}
D_{P_\mathrm{out}}=\frac{1}{2\period}\int_0^\period\int_0^\period\dd{\tau}\dd{\tau'}\expval{x^2(\tau)x^2(\tau')}\dot{\lambda}(\tau)\dot{\lambda}(\tau')/4-\frac{\period}{2}P_\mathrm{out}^2.
\end{equation}
The the correlation function $\expval{x^2(\tau)x^2(\tau')}$ can be written as
\begin{equation}
\expval{x^2(\tau)x^2(\tau')} = \int_{-\infty}^\infty\int_{-\infty}^\infty\dd{x}\dd{x'} x^2 x'^2 p(x,x',\tau,\tau'),
\label{eq:example:D_integrand}
\end{equation}
where $x\equiv x(\tau)$, $x'\equiv x'(\tau)$ and $p(x,x',t,t')$ is the joint probability density.
Here, $x$ and $x'$ are Gaussian distributed according to~\eqref{eq:examples:ps_x} .
Due to the linearity of the Langevin equation, every linear combination of $x$ and $x'$ is also Gaussian distributed.
Hence, the joint probability density $p(x,x',\tau,\tau')$ is a bivariate normal distribution and the expectation value in~\eqref{eq:example:D_integrand} is consequently given by
\begin{equation}
\expval{x^2(\tau)x^2(\tau')} = \expval{x^2(\tau)} \expval{x^2(\tau')} + 2\expval{x(\tau)x(\tau')}^2.
\label{eq:examples:correlation_func}
\end{equation}
By solving the Langevin equation, we can evaluate the correlation function \eqref{eq:examples:correlation_func} and obtain the following expression for the diffusion coefficient
\begin{equation}
D_{P_\mathrm{out}}=\frac{1}{\period}\int_0^\period\int_0^\tau\dd{\tau}\dd{\tau'}\frac{1}{2}\dot{\lambda}(\tau)\dot{\lambda}(\tau')e^{2(I(\tau')-I(\tau))}
\label{eq:example:D_power}
\end{equation}
with
\begin{equation}
I(\tau)\equiv \mu\int_0^\tau\dd{t'}\lambda(t').
\end{equation}
\bibliographystyle{apsrev4-1}
\bibliography{./refs.bib}